\documentclass[a4paper]{article} 

\usepackage{graphicx} 
\usepackage[caption=false]{subfig}
\usepackage[font=small,labelfont=bf]{caption}
\usepackage{hyperref} 
\usepackage{tabularx} 
\usepackage{color} 
\usepackage{amsmath} 
\usepackage{amssymb} 
\usepackage{amsfonts} 
\usepackage{amsxtra} 
\usepackage{upgreek} 
\usepackage{enumerate} 
\usepackage{authblk}
\usepackage{pdflscape}
\usepackage{array} 
\newcolumntype{x}[1]{>{\centering\arraybackslash}p{#1}}

\newcommand*{\matvec}[1]{\mathbf{#1}}
\newcommand*{\cov}{\mathrm{cov}}
\newcommand*{\cor}{\mathrm{cor}}

\graphicspath{{./figs/}}

\usepackage{fancyhdr}
\usepackage{currfile}
\title{Applications of robust estimators of covariance in examination of inter-laboratory study data}

\author[1]{Stephen L R Ellison}
\affil[1]{LGC Limited, Queens Road, Teddington, Middlesex TW11 0LY, UK}

\begin{document}

\maketitle

\begin{abstract}
This paper illustrates the use of selected robust estimators of covariance or correlation in the identification of anomalous laboratory results in inter-laboratory data. It is shown that robust estimators can substantially reduce the impact of outlying values on multivariate confidence regions and consequently lead to sharper identification of anomalies, even where traditional outlier detection may fail to locate anomalous results.
\end{abstract}

\section{Introduction}

Many inter-laboratory studies involve the collection of results for more than one measurand from each participant. For example, in a collaborative study aimed at validation of a new standard measurement procedure, results for more than one test material are usually collected, either to obtain information on precision at different levels or as part of a split-level design \cite{aoac75}. In reference material certification by inter-laboratory study, results for multiple analytes in the same candidate reference material may be obtained, or a separate quality control material of known properties may be included for assessing laboratory performance \cite{G35}. Proficiency testing rounds also frequently collect data for multiple measurands, on multiple test items, or both. In these circumstances, the identification of anomalous results can be challenging, as it is possible for a laboratory to have results with an acceptable range for each individual measurand on each test item but nonetheless differ substantially from the remainder of the population in terms of the general pattern of results. Identifying such anomalies is an important step in initial data inspection and `clean-up'.

The problem of location outliers in interlaboratory studies is well known \cite{HUND2000145}.  Standards for interlaboratory study, such as ISO 5725 \cite{ISO5725-2} and the IUPAC protocol \cite{IUPAC-collab-1994} for collaborative study of test methods, routinely include procedures for detection of location outliers. Similarly, proficiency testing guidance such as ISO 13528 \cite{ISO13528-2015} includes outlier inspection and accommodation methods. While most early guidance used univariate outlier detection, robust statistical methods, as suggested by the Royal Society of Chemistry's Analytical methods Committee \cite{AMC89-1}, have become a widely used alternative for accommodation of outlying results \cite{HUND2000145}. So far, however, essentially all of these rely on application of univariate methods, applied to results for one measurand or test material at a time. These methods do not take account of the frequent strong correlation visible in many interlaboratory studies, and can miss unusual patterns of results. This multivariate problem suggests a multivariate approach. 

Several approaches for outlier identification in multivariate data are available. For bivariate data, Youden plots can be effective \cite{aoac75}. For multivariate data, principal component analysis \cite{chatcol80} and measures such as Mahalanobis distance \cite{mahal1936} can be valuable aids. However, while visual inspection is always useful for inspection, it is often useful to include criteria for declaring an observation as anomalous. For example, for univariate outlier detection, critical values for common statistical outlier tests are used to decide whether follow-up action is appropriate. For multivariate data, such criteria will commonly require information on covariance between different measurands. Since most interlaboratory data sets show at least some outliers, however, the usual covariance measures, such as sample covariance and Pearson correlation (usually denoted \textit{r}), can badly overestimate the covariance of the underlying distribution. What is needed are methods of estimating covariance that are robust to the presence of outlying values.

There are now a number of outlier-resistant procedures for obtaining estimates of covariance or correlation. These include, for example, a simple pairwise procedure due to Gnanadesikan and Kettenring \cite{GK72}, and more complex iterative procedures such as the minimum covariance determinant method \cite{fastMCD} or the ``OGK'' estimator \cite{OGK02}. Rank correlation (either Spearman or Kendall \cite{kendall38}) is also relatively resistant to extreme values compared to the usual Pearson correlation. Despite their availability, these have rarely been applied to the analysis of interlaboratory study data. Lischer noted the possible utility of a robust Mahalanobis score \cite{Lischer1996}, but considered available robust covariance estimators insufficiently reliable for the purpose at that time. Dueck and Lohr \cite{dueckLohr2005} later proposed a procedure based on M-estimation and demonstrated its use for outlier identification in a biological measurement study with two measurements per test item. ISO 13528:2005 \cite{ISO13528-2005} suggested the use of methods based on rank correlation to support interpretation of Youden plots, but did not include any formally robust statistical methods. There seem to be no examples of robust covariance use in inter-laboratory studies for analytical chemistry.

This paper is accordingly intended to illustrate the use of selected robust estimators of covariance or correlation in the identification of anomalous laboratory results in inter-laboratory data, focussing on the situation of a single laboratory result (or mean) for each test item and measurand. The discussion begins with a brief overview of the concept of covariance and a summary of some important characteristics of robust estimators. Selected robust estimators of covariance are then described briefly, before illustrating their application to review of interlaboratory study data. 

\section{Covariance and correlation}

Covariance is commonly taken as a property of a pair of variables or data sets; the classical covariance estimator for vectors $\matvec{x}_{i}=(x_{1i}, x_{2i}, \ldots, x_{ni})$, $\matvec{x}_{j}=(x_{1j}, x_{2j}, {\ldots}, x_{nj} )$ of length $n$ can be written

\begin{equation}
\label{ref-001}
\cov\left(\matvec{x}_{i},\matvec{x}_{j}\right)=\frac{\sum _{k=n}^{1}\left(x_{ik}- \overline{\matvec{x}}_{i}\right)\left(x_{jk}- \overline{\matvec{x}}_{j}\right)}{n- 1}
\end{equation}
and the correlation coefficient $r_{ij}$ is 

\begin{equation}
\label{ref-002}
r_{ij}=\frac{\mathrm{cov}\left(\matvec{x}_i,\matvec{x}_j\right)}{s(\matvec{x}_i)s(\matvec{x}_j)}
\end{equation}

where $s(\matvec{x}_{i})$, $s(\matvec{x}_{j})$ denote the respective sample standard deviations. Note that when $j=i$, $\mathrm{cov}(\matvec{x}_{i}, \matvec{x}_{j})$ is just the variance for $\matvec{x}_{i}$. While variance is always positive, a valid covariance can be positive or negative; its magnitude is also less than or equal to the product of the two standard deviations. A valid correlation coefficient can therefore be in the interval $[-1, 1]$. These properties are guaranteed by equations \eqref{ref-001}  and \eqref{ref-002}  but, as will be seen below, not necessarily guaranteed by other possible estimators. For multivariate data with \textit{p} variables, the classical covariance is described by a covariance matrix \textbf{V}, the elements of which are given by

\begin{equation}
\label{eqn:covmat}
\mathbf{V}_{i,j}=\mathbf{V}_{j,i}=\mathrm{cov}\left(\matvec{x}_{i},\matvec{x}_{j}\right);\, \, \, i,j=1,2, \, \ldots, \, p
\end{equation}

While this is simply a symmetric square matrix containing the variances along the leading diagonal and the pairwise covariances elsewhere, a less obvious but important additional feature is that a valid covariance matrix is, (for \textit{p~{\textless}~n})\textit{,} positive definite; that is, its determinant is strictly greater than zero. For \textit{p~}${\geq}$~$n$, however, the covariance matrix is singular and has zero determinant. For multivariate (rather than pairwise) applications, this can have additional implications for the choice of robust covariance estimator. 

\section{Robust covariance estimators}

\subsection{Performance characteristics for robust estimators}

Any choice of robust estimator is influenced by a range of considerations, so it is useful to summarise some of the main performance characteristics of robust estimators. The most important statistical properties are usually breakdown point, efficiency and bias. The breakdown point is a measure of the proportion of outlying values that can be tolerated. There are at least two different definitions in univariate data; for the present paper we take breakdown point as the proportion of values that can move to ${+\infty}$ without the estimate becoming infinite. The efficiency, $e$, is usually taken as the ratio of the classical variance to the variance of the estimator in question for normally distributed data; a practical way of looking at efficiency is that $n/e$ is the number of observations needed to give the same uncertainty as the corresponding classical estimator applied to $n$ data. Efficiency is often much more important for location estimators (functions that give an estimate of mean value) as it directly affects the uncertainty of estimates. Bias indicates whether an estimator, on average, over- or under-estimates the corresponding population parameter. Ideally, one seeks an unbiased estimator; where this is impractical, estimators may be adjusted for \textit{consistency}; that is, while they may be biased for small $n$, as $n$ tends to infinity, the expectation of the estimator tends to the population value. 

\begin{figure}
	\centering
	\includegraphics[width=0.6\textwidth]{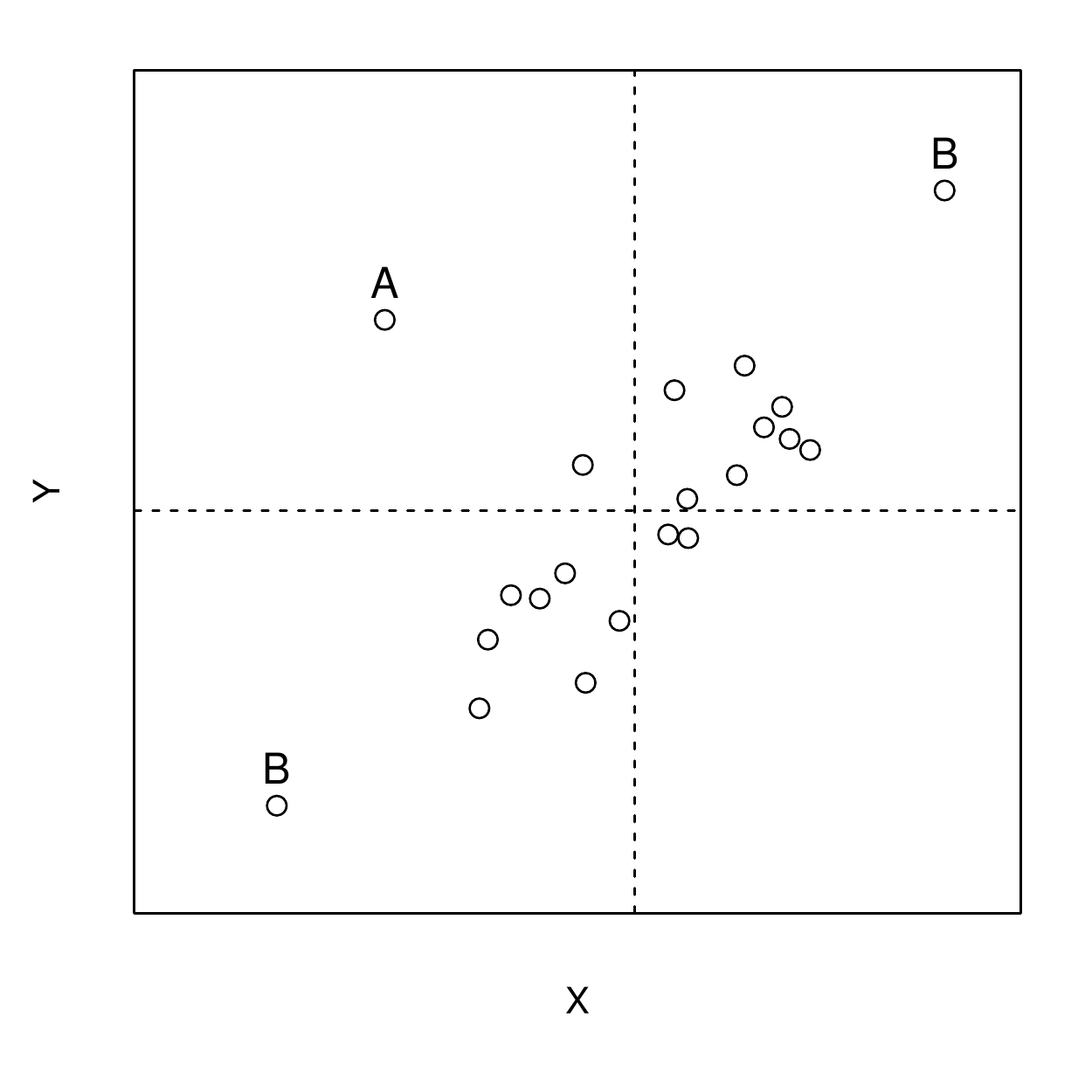}
	\caption{Two different kinds of outlier in bivariate data. A: ``correlation outlier''; B, ``tail outliers''. Vertical and horizontal dashed lines are through the medians of the data, for visual guidance.}

	\label{ref-032}
\end{figure}

A particular issue in assessing the performance of robust estimators in the multivariate case is that the nature of outliers is important. Figure~\ref{ref-032} shows two important cases. ``Tail outliers'' (B in Figure~\ref{ref-032}) are just the familiar univariate outliers in the same direction for both (or all) sets of data. Another type, labelled A in Figure~\ref{ref-032}, arise from a population with correlation ${-\rho}$, when the bulk of the data have correlation ${\rho}$. These have been termed ``correlation outliers'' \cite{Croux2010}. These typically decrease correlation unless very extreme. Unless extreme, they can not easily be detected by univariate data inspection; note that point A in Figure~\ref{ref-032} is not particularly extreme for either variable. Resistance to correlation outliers is a desirable property of robust covariance estimators; assessment of bias accordingly needs to consider both  ``tail'' and ``correlation'' outliers.

Two other practical characteristics can also be important; computational complexity and computing time. Speed and simplicity can -- particularly for inspection and review -- be more important than the best possible efficiency or bias.

We now present a number of established robust covariance estimators and review them against these criteria.

\section{Some outlier-resistant estimators of covariance or correlation}

\subsection{Rank correlation}

There are two common measures of rank correlation; Spearman's ${\rho}$ and Kendall's ${\tau}$ \cite{kendall38}. Spearman's ${\rho}$ for two sets of data ($\matvec{x}$, $\matvec{y}$) is simply the correlation coefficient \eqref{ref-002} calculated from the ranks $\left(R(x_{i}), R(y_{i})\right)$ within each set, that is, the numerical location of each data point when the set is sorted into ascending order. Spearman's ${\rho}$ can also be calculated from 

\begin{equation}
\label{eqn:rho-alt}
	\rho =1- \frac{6\sum _{i=1,n}d_{i}^{2}}{n(n^{2}- 1)}
\end{equation}

where $d_{i}=R(x_{i})-R(y_{i})$ and $n$ is the number of (bivariate) data points.

Kendall's ${\tau}$ relies on the numbers $N_{c}$ and $N_{d}$ of, respectively, ``concordant'' and ``discordant'' points in a multiple pairwise comparison of the data sets. A pair of points $\left(x_{i}, y_{i}\right)$ and $\left(x_{j}, y_{j}\right)$ is considered concordant if ($x_{j}$-$x_{i}$) and $\left(y_{j}-y_{i}\right)$ have the same sign; discordant if they have opposite sign, and is not counted if $(x_{j}=x_{i})$ or ($y_{j}=y_{i})$. Then, Kendall's ${\tau}$ is given by

\begin{equation}
\label{eqn:kendall-tau}
\tau =\frac{N_{\mathrm{c}}- N_{\mathrm{d}}}{N_{\mathrm{c}}+N_{\mathrm{d}}}
\end{equation}

Like the correlation coefficient \textit{r}, these measures are both limited to the range {$[-1,1]$}.  Both rank correlations are quite efficient, with efficiencies above 70\% at the normal; for both, the asymptotic efficiency is lowest near ${\rho}=1$, increasing to about 90\% as ${\rho}$ tends to zero. \cite{Croux2010} 

Because ranks are themselves insensitive to distribution, measures of correlation based on ranks already provide fair resistance to outlying values. Croux and Delon found good resistance for the Spearman and Kendall correlation at 1\% contamination by ``worst-case'' outliers (chosen for their extreme effect on rank correlation); for both, effects became appreciable at 5\% contamination, though Kendall's ${\tau}$ showed smaller effects. \cite{Croux2010} For ``correlation outliers'', both of these correlation measures tolerate 5-10\% outlier contamination well, though with a bias towards smaller correlation. 

Both $\rho$ and $\tau$ can be used to construct covariance estimates by multiplication by the product $s(\matvec{x})s(\matvec{y})$ of the respective standard deviations, using the relationship in eq. \eqref{ref-002}. For outlier resistance for the corresponding covariance estimates, the standard deviations can be replaced by robust standard deviations for each of the two variables concerned, giving, for example: 

\begin{equation}
\label{eqn:covRS}
\mathrm{cov}_{RS}\left(\matvec{x},\matvec{y}\right)=\rho s^{*}(\matvec{x})s^{*}(\matvec{y})
\end{equation}

where $s^{\mathrm{*}}(.)$ denotes a robust estimate of standard deviation and the subscript \textit{RS} indicates a \underline{R}ank correlation using \underline{S}pearman's method. The combination of a rank correlation with robust estimates of standard deviation provides considerably improved outlier resistance. For this reason, ISO 13528:2005 \cite{ISO13528-2005} recommended the use of Spearman's rank correlation, supplemented by the ``Algorithm A'' robust estimates of standard deviation, as a basis for confidence ellipsoids in Youden plots \cite{ISO13528-2005}. The efficiency and breakdown for covariance estimators such as \eqref{eqn:covRS} is dictated primarily by the robust standard deviation estimator; for example, use of the scaled median absolute deviation (MAD$_{\mathrm{e}}$, \cite{AMC89-1}) provides 50\% breakdown but efficiency at the normal of only 37\% \cite{MaronnaYohai}, while Rousseeuw's $Q_{\mathrm{n}}$ retains the high  breakdown of 50\% but has efficiency of 82\% \cite{Qn1993}.

\subsection{Gnanadesikan-Kettenring pairwise estimator (GK)}

Gnanadesikan and Kettenring \cite{GK72} noted the identity 
\begin{equation}
\mathrm{cov}\left(\mathbf{x,y}\right)=\left[\mathrm{var}\left(\matvec{x}\mathbf{+}\matvec{y}\right)- \mathrm{var}\left(\matvec{x}\mathbf{-}\matvec{y}\right)\right]/4
\end{equation}
and proposed the simple covariance estimate

\begin{equation}
\label{ref-004}
\mathrm{cov}_{GK}\left(\matvec{x},\matvec{y}\right)=\left(s_{1}^{*2}- s_{2}^{*2}\right)/4 
\end{equation}
where $s_{1}^{*}$ and $s_{2}^{*}$ are robust standard deviations for $\left(\matvec{x}+\matvec{y}\right)$ and $\left(\matvec{x}- \matvec{y}\right)$ respectively. This gives a very simple estimator, as it relies only on the availability of a robust estimator of standard deviation. The principle is also very flexible, in that it can use any robust estimate of scale, including very simple estimators such as MAD$_{\mathrm{e}}$ and more efficient estimators such as Rousseeuw's $Q_{\mathrm{n}}$ \cite{Qn1993}. The breakdown and efficiency follow the properties of the robust scale estimator used. It does, however, suffer from the disadvantage that the magnitude of a calculated $\cov_{GK}$ is not guaranteed to be smaller than the product of the robust standard deviations; essentially because location-dependent weights used in calculating $s_{1}^{*}$ and $s_{2}^{*}$ are generally not the same as the weights used for corresponding data points in calculating $s^{*}\left(\matvec{x}\right)$ and $s^{*}\left(\matvec{y}\right)$. A covariance matrix constructed in this way -- even for two variables -- is therefore not guaranteed to be positive definite 
\cite{GK72, OGK02}.

\subsection{Gnanadesikan-Kettenring estimate of correlation (RGK)}

Since $\cov_{GK}$ can lead to invalid covariance estimates, and therefore estimates of correlation outside the admissible range $[-1, 1]$, Gnanadesikan and Kettenring proposed an alternative robust estimator of correlation, here denoted $\cor_{GK}$, which is guaranteed to be in the admissible range. $\cor_{GK}$, for data $\mathbf{x, y}$,  is constructed as follows:

\begin{enumerate}[i)]
	\item Calculate scaled data $\mathbf{z}_x = \matvec{x}/s^{\mathrm{*}}(\matvec{x})$,   $\mathbf{z}_y = \matvec{y}/s^{\mathrm{*}}(\matvec{y})$, where $s^{\mathrm{*}}(.)$ denotes a robust estimate of standard deviation, as before; 
	
	\item Calculate robust standard deviations $s_{z+}^{\mathrm{*}}$, $s_{z-}^{\mathrm{*}}$ of $\mathbf{z}_x + \mathbf{z}_y$ and $\mathbf{z}_x - \mathbf{z}_y$  respectively;
	
	\item $\cor_{GK}$ is then calculated as
	\begin{equation}
	\label{eqn:corGK}
	\cor_{GK}(\matvec{x, y}) = \frac{s_{z+}^{\mathrm{*}2} - s_{z-}^{\mathrm{*}2}}{s_{z+}^{\mathrm{*}2} + s_{z-}^{\mathrm{*}2}}
	\end{equation}
\end{enumerate}

Since equation \eqref{eqn:corGK} is a difference of (robust) variances divided by their sum, it must always be in the range $[-1, 1]$.

Given $\cor_{GK}$, it is now possible to re-use the original scale estimates $s^{\mathrm{*}}(\matvec{x})$ and $s^{\mathrm{*}}(\matvec{y})$ in the same way as equation \eqref{eqn:covRS} to obtain the covariance estimator:
\begin{equation}
\label{eqn:covRGK}
\cov_{RGK} = \cor_{GK}(\matvec{x, y}){\mathrm{*}}s^(\matvec{x})s^{\mathrm{*}}(\matvec{y})
\end{equation}
where the subscript ``RGK'' denotes that the covariance esimator arises from the Gnanadesikan-Kettenring estimate of correlation $\cor_{GK}$.

Like $\cov_{GK}$, $\cov_{RGK}$ only requires a means of calculating a robust standard deviation; one suggested implementation \cite{Shevlyakov_Smirnov_2016, ShevPas1987} used the simple and well-known $\mathrm{MAD}$ estimator (which does not need scaling for consistency as the scaling factor cancels in \eqref{eqn:corGK}) to form a ``MAD correlation coefficient''. 

As for $\cov_{GK}$, the efficiency and breakdown properties for $\cov_{RGK}$ broadly follow those of the robust standard deviation estimator $s^*$ used in \eqref{eqn:covRGK}.
 
\subsection{Orthogonalized Gnanadesikan-Kettenring estimator \newline (OGK)}

To guarantee a positive definite covariance matrix, Maronna and Zamar \cite{OGK02} proposed an iterative extension of $\cov_{GK}$, denoted $\cov_{OGK}$. The algorithm involves iteratively scaling the variables; constructing an initial correlation matrix $\matvec{U}$ by applying $\cov_{GK}$ (equation~\eqref{ref-004}) to the scaled variables; extraction of the eigenvalues \textbf{e}$_{i}$ of $\matvec{U}$; and use of these and the initial robust standard deviations to form a new covariance estimate. As a side effect of a re-weighting step in the algorithm, Mahalanobis distances are also calculated and available. A full description is beyond the scope of the present paper, but implementations are readily available; in particular in the R {\texttt robustbase} package \cite{robustbase} and the R {\texttt rrcov} package \cite{rrcov-article, rrcov} which, like R itself, are free to use. $\cov_{OGK}$ does require specification of a robust standard deviation (scale) estimator; Maronna and Zamar \cite{OGK02} preferred a ``$\tau$ scale'' estimate proposed by Yohai and Zamar \cite{YohaiZamar1988}, for speed and because it gave better results than ${\mathrm{MAD_e}}$ in simulations. Although computationally more complex, $\cov_{OGK}$ is relatively fast, even for relatively high-dimensional data, and successfully guarantees a valid robust estimate of covariance. Breakdown and efficiency again largely follow the particular estimators chosen in the intermediate $\cov_{GK}$ calculations.

\subsection{Minimum Covariance Determinant (MCD) estimator}

\begin{figure}
	\centering
	\subfloat[]{
		\includegraphics[width=0.4\textwidth]{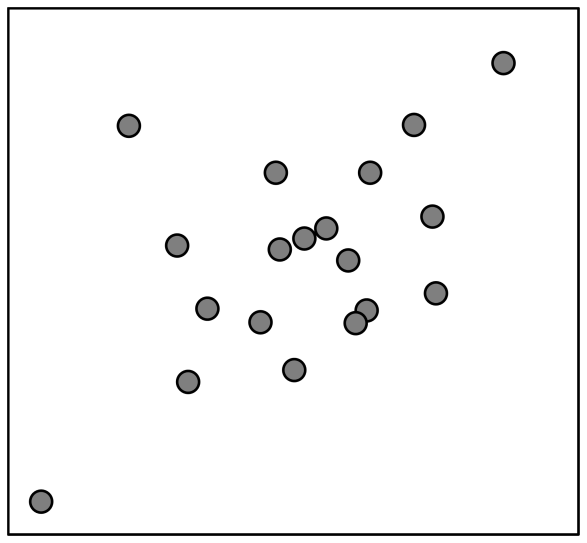}
		\label{fig:mcd-a}
	} 
	\subfloat[]{
		\includegraphics[width=0.4\textwidth]{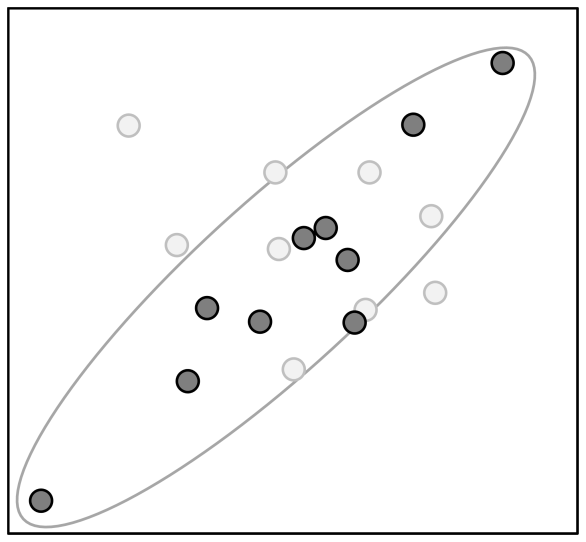}
		\label{fig:mcd-b}
	}
	
	\subfloat[]{
		\includegraphics[width=0.4\textwidth]{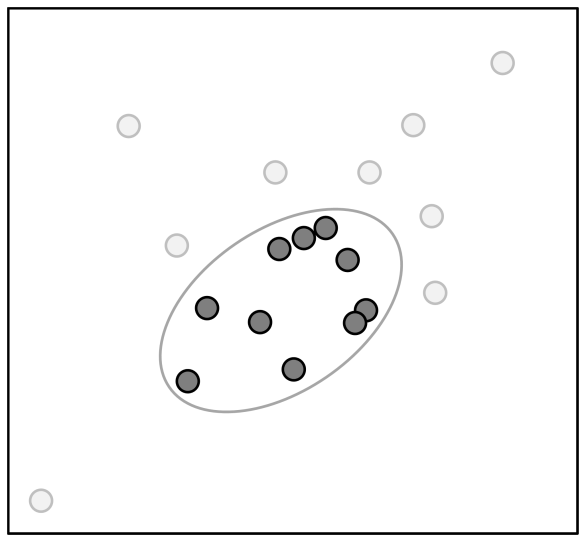}
		\label{fig:mcd-c}
	}
 
	\caption{Schematic of MCD estimation. The minimum covariance determinant estimate is formed from a multivariate data set  \protect\subref{fig:mcd-a} of $n$ data points by \protect\subref{fig:mcd-b} selecting a subset of size \textit{h} (dark grey in \protect\subref{fig:mcd-b}), determining the covariance of the subset, taking the determinant of that covariance matrix, and continuing iteratively to find the subset with the smallest such determinant \protect\subref{fig:mcd-c}. Having found the relevant subset, the covariance is determined and re-scaled, typically to return an unbiased estimate of covariance for multivariate normal data.}
	\label{ref-033}	
\end{figure}

An alternative strategy can be thought of as a multidimensional equivalent of data set truncation to remove extreme values before calculating the usual covariance and then re-scaling, in a manner reminiscent of the use of a trimmed mean or Shorth (``shortest half'', \cite{andrews72}) estimate. An efficient example is the minimum covariance determinant (MCD) estimator proposed by Rousseeuw {\cite{rousseeuw1985-mcd}}. The method finds the subset of $h~(n/2\leq h \le n)$ data points in a set of $n$ that has the classical covariance matrix with the smallest determinant. The process is illustrated in Figure~\ref{ref-033}. Since there are generally $n!/\left[h!\left(n- h\right)!\right]$ subsets of size \textit{h}, this can be slow. A fast algorithm (``FastMCD'' or FMCD) is, however, available \cite{fastMCD} and has been applied successfully to data sets of the order of 10~000 data points; far above the size of typical inter-laboratory data sets. Implementations are available in, for example, the R \texttt{robustbase} \cite{robustbase} and \texttt{rrcov} \cite{rrcov, rrcov-article} packages, and the R \texttt{MASS} package \cite{MASS-book}. The estimator, which we denote $\cov_{MCD}$, has the advantage that the covariance matrix is guaranteed to be positive definite as long as the number of variables is less than $h$, and the centroid of the chosen subset additionally provides a robust estimate of location. The breakdown point, $h/n$, can be as good as 0.5 (as robust as the median) depending on the choice of \textit{h}. Common recommendations are $h=0.5n$ and $h=0.75n$. Efficiency at the normal for MCD (with $h=0.75n$ and reweighting as recommended by Rousseeuw \cite{rousseeuw1985-mcd}) is good, at between about 60\% and 90\% depending on number of variables; lower dimensionality is associated with lower efficiency \cite{CroukHaes1990}. 

A practical disadvantage of the MCD estimator is that the number of variables must be less than \textit{h}, or the determinant vanishes for any subset. Although this is not often a problem for bivariate covariance estimation, it can appreciably restrict the application of MCD in inter-laboratory studies of modest size with many measurands. A further practical difficulty -- other than the computational complexity of the FastMCD algorithm -- is that, in our hands, some implementations appeared to be biased low for under about 60 data points, ruling out their use for typical reference material certification studies. Some care must accordingly be taken in choosing the implementation. Of the MCD implementations examined, the implementation in \cite{robustbase} was found to give good results for the modest data set size in the examples shown here.

\section{Experimental}

\subsection{Illustrative data}

To illustrate some applications of robust estimators of covariance, we use two inter-laboratory data sets from a reference material certification exercise, used to certify a drinking water material. Two materials were circulated; a candidate reference material (``RM'') and a performance control material (``QC''). The candidate reference material was a drinking water from a domestic supply, fortified for some elements of interest. The performance control material was made up to similar (but intentionally not identical) levels to the candidate RM by spiking demineralised water. Laboratories measured both materials, results for the QC material being used to check for serious measurement bias. Five replicates were run for the candidate RM; three for the QC material. Summary data (laboratory means) are given in {\hyperref[ref-038]{Table~S1}} and {\hyperref[ref-039]{Table~S2}} (provided as supplementary material). {\hyperref[ref-038]{Table~S1}} gives mean results for potassium, for both materials. {\hyperref[ref-039]{Table~S2}} gives summary results for eight elements for the candidate reference material only; the mean of five replicates is given. 

\subsection{Computational methods}

All calculations and plots used R, version 3.4.4 \cite{Rstats}. Youden plots with robust confidence regions were produced using the \texttt{metRology} package for R, version 0.9-28 \cite{metRology}, supported by the \texttt{robustbase} package \cite{robustbase}. Robust PCA used the \texttt{rrcov} package \cite{rrcov}. All computations were performed on an Intel-based PC running an Intel Pentium G3528 running at 3.2~GHz; the operating system was Microsoft Windows 10 Pro.

\section{Applications}

\subsection{Robust confidence regions for Youden plots.}

Youden plots \cite{aoac75} are a convenient graphical means of checking for an important between-laboratory effect in an interlaboratory study. They consist of a scatter plot of laboratory mean observations on two materials. If there is no laboratory effect, these should form an uncorrelated scatter; an important laboratory effect (comparable to or greater than the within-laboratory standard deviation) leads to marked correlation. Although originally proposed for review of data from a split-level design, in which the two materials have very similar concentrations of a given analyte, the plot also works well in most circumstances where two materials are measured using the same procedure and where multiple measurands are determined simultaneously for the same material (for example, simultaneous measurement of multiple elements by ICP-MS). Because of their convenience and simplicity, they are also suggested for review of proficiency testing data \cite{ISO13528-2015}. To confirm whether individual data points can be considered anomalous, the 2005 edition of ISO 13528 additionally suggested the use of confidence ellipses based on rank correlation, on the (obsolete) grounds that ``There is a need for a robust method of calculating the ellipse, but the details of such a method have not yet been worked out.'' (ref. \cite{ISO13528-2005}, sec 8.5.2.1 Note 2).

In this RM study, a QC and candidate RM were circulated and measured. Youden plots of the candidate RM result against the corresponding QC material data were appropriate for inspection. 

\begin{figure}
	\centering
	\includegraphics[width=0.8\textwidth]{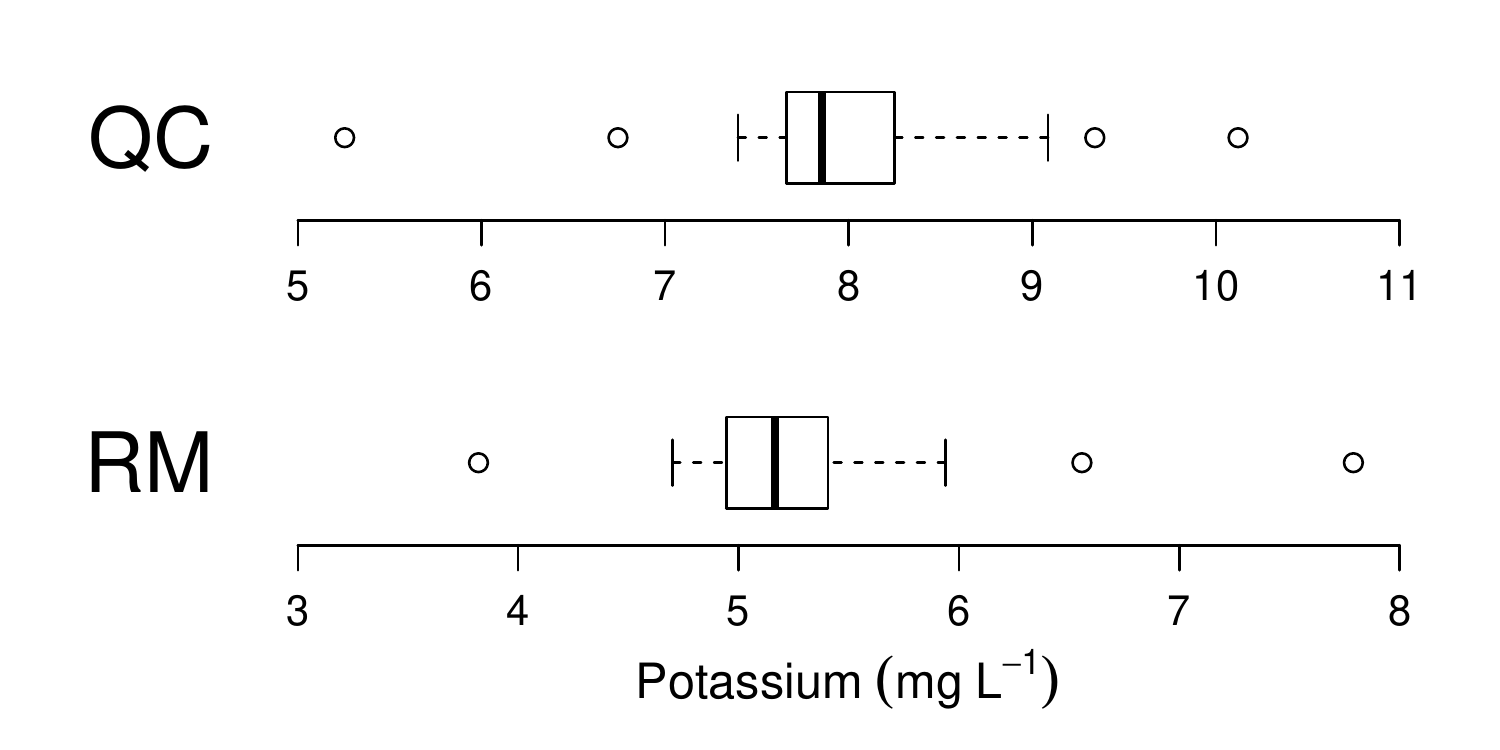}
	\caption{Box plots of mean results for potassium from 25 laboratories for a candidate RM (``RM'') and a performance control material (``QC'') in a reference material characterisation exercise.}

	\label{ref-034}
\end{figure}

Before considering Youden plots, it is useful to look at the data from a univariate perspective. Figure~\ref{ref-034} shows data for potassium in both materials. While it is clear that there are several tail outliers, there is no immediate indication of any other anomaly. 

\begin{figure}[!h]
	\centering
	\subfloat{\includegraphics[width=0.47\textwidth]{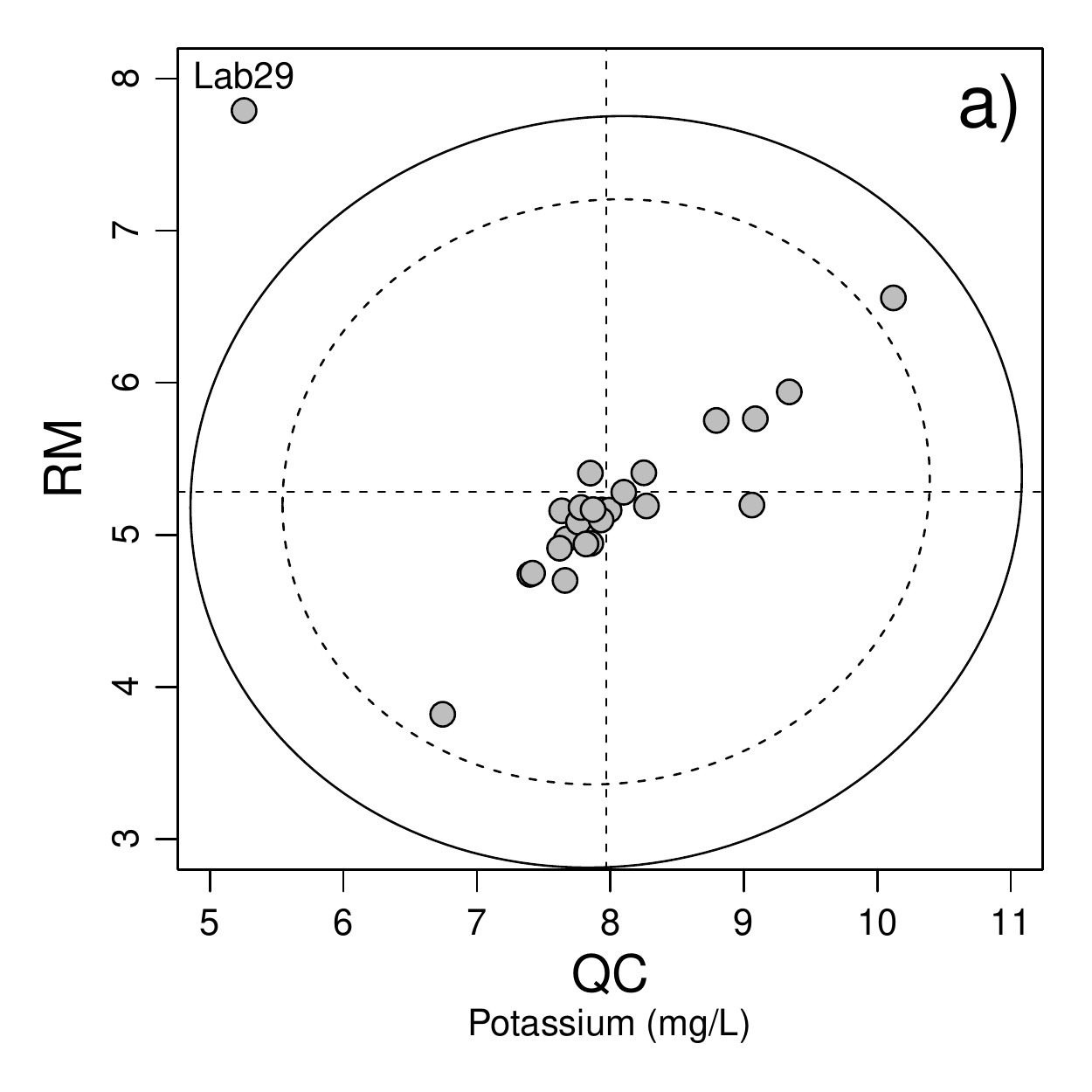} \label{fig:youden-a}}
	\quad
	\subfloat{\includegraphics[width=0.47\textwidth]{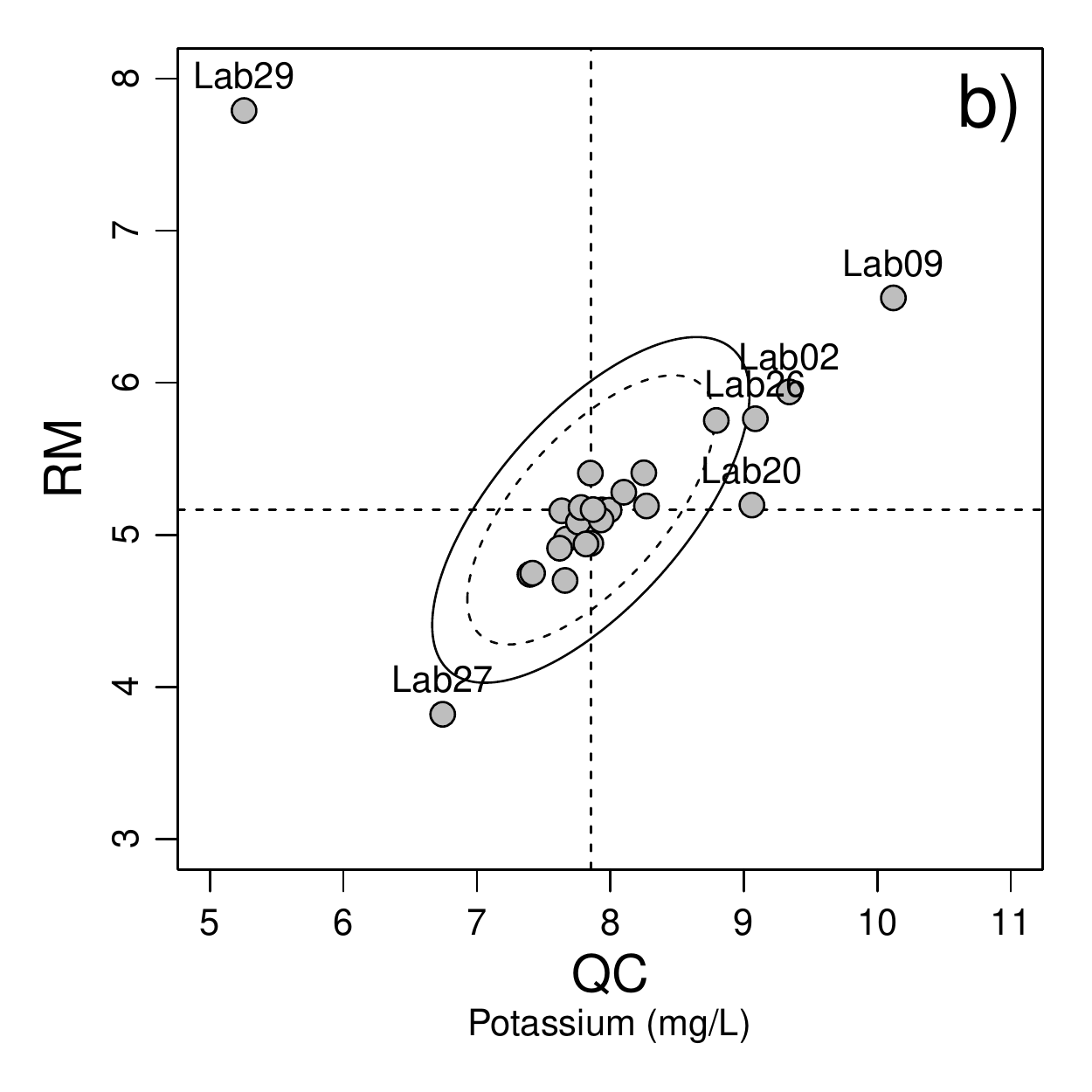} \label{fig:youden-b}}
	
	\subfloat{\includegraphics[width=0.47\textwidth]{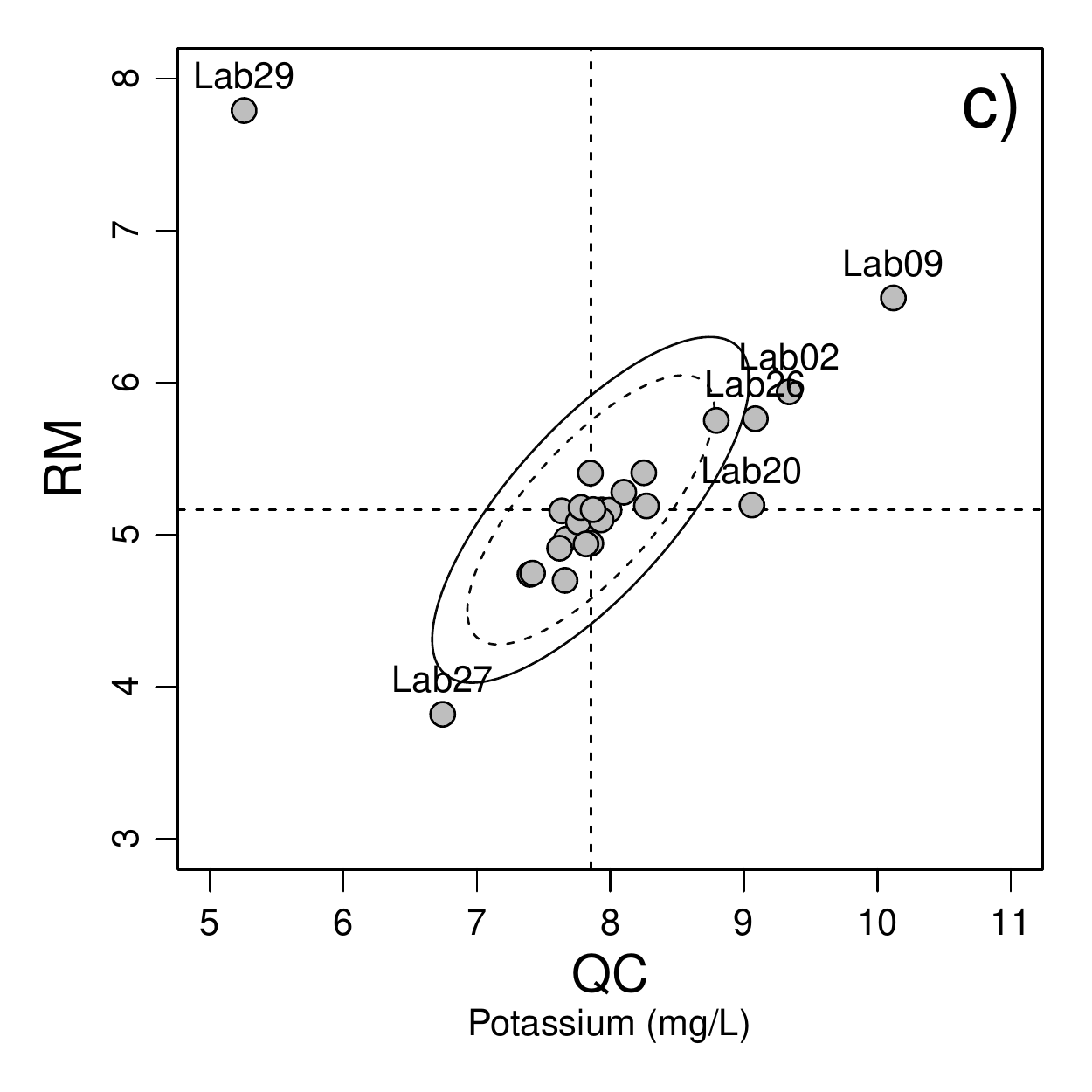} \label{fig:youden-c}} 
	\quad
	\subfloat{\includegraphics[width=0.47\textwidth]{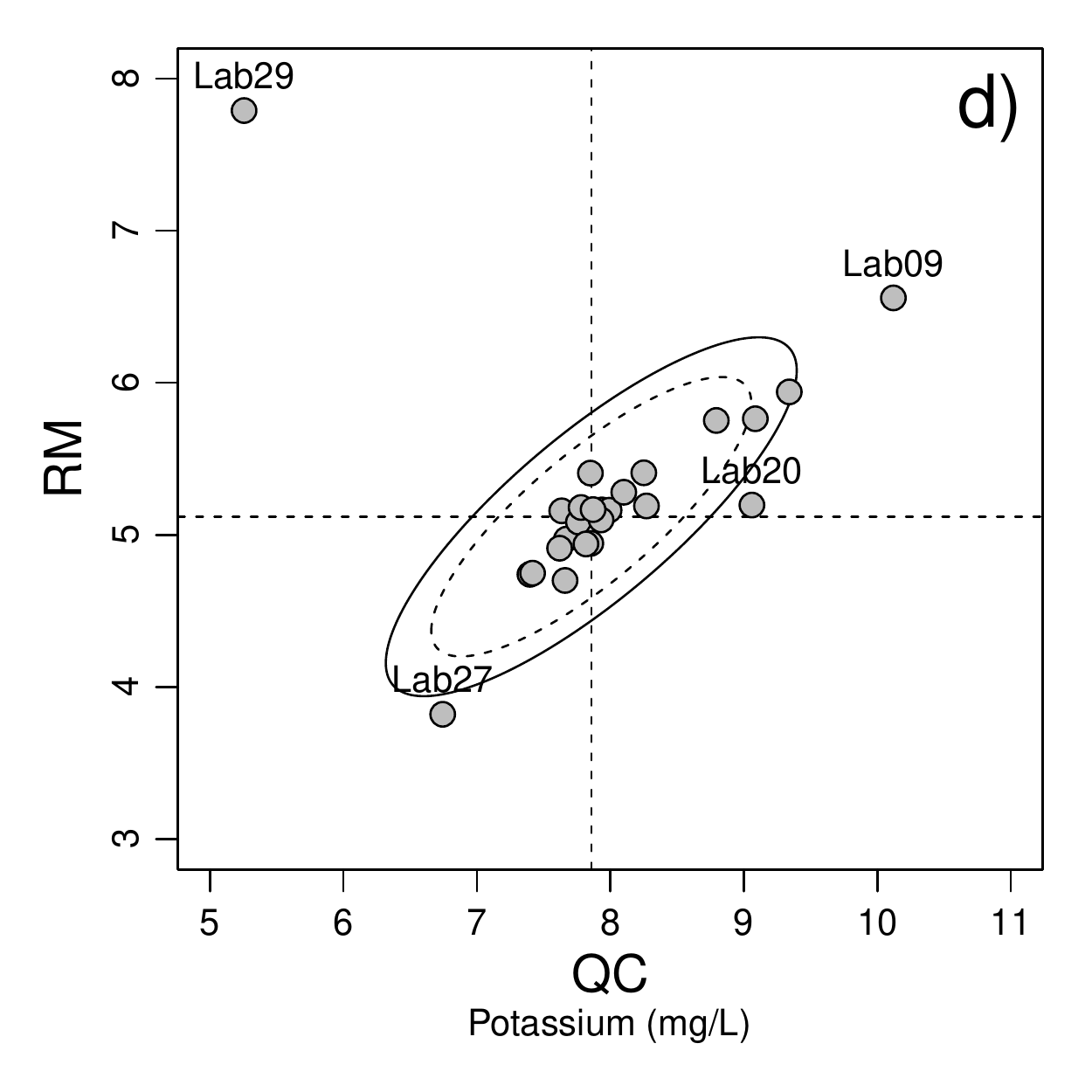} \label{fig:youden-d}} 
	
	\caption{Youden plots of laboratory means for potassium on two materials, with data confidence ellipses at 95\% (dashed line) and 99\% (solid line). The estimators for the covariance were a) classical covariance (equation \eqref{ref-001}; b) Spearman rank correlation with $\mathrm{MAD_e}$ scale; c) Gnanadesikan-Kettenring correlation estimator ($\cov_{RGK}$) with $\mathrm{MAD_e}$ scale; d) orthogonalised Gnanadesikan-Kettenring estimator ($\cov_{OGK}$). Vertical and horizontal lines are drawn through the centroid as calculated by the relevant classical or robust location estimate. Laboratory identifiers are shown for any observation outside the 99\% confidence region.}

	\label{ref-035}
\end{figure}

Youden plots add a new insight and illustrate the advantage of robust estimates of covariance. Youden plots for the potassium data, with confidence ellipses calculated by four covariance methods, are given in Figure~\ref{ref-035}. In all cases, the ellipses are calculated from the respective covariance estimate using the methods described by Jackson \cite{jacksonIQC, jacksonTech59}; a summary is given in the Annex. Note that these ellipses are constructed to include a given proportion of observations, and not the confidence region for the centroid.

It is immediately clear from the Youden plots that one laboratory (Lab29) forms a severe correlation outlier, visible at the top left of each plot. This almost certainly arises from confusion in sample labelling, as the values reported would not have been unreasonable for the alternate materials and the laboratory was found to be a similar off-diagonal outlier for several other elements, with the extent of the discrepancy increasing with the difference in concentrations for the two materials.

For the purpose of the present paper, however, the effect of the different covariance estimators is of most interest. The ellipse based on classical covariance, denoted a) in the Figure, is clearly substantially inflated by all the outliers. It is also clear from the near-circular confidence ellipse that the correlation (0.04) has been severely reduced by the correlation outlier. Both effects lead to a considerably inflated confidence region. By comparison, all of the robust estimators give much smaller confidence regions and show substantially greater correlation; for this data set, the correlations from Spearman, RGK, OGK and MCD estimators are 0.67, 0.75, 0.81 and 0.86 respectively. For comparison, omitting Lab29 gives a classical correlation of 0.91, in part due to the two marked tail outliers but clearly showing the drastic effect of the off-diagonal observation. 

There is less variation among the confidence regions calculated from robust covariance estimators. The Spearman-based $\cov_{RS}$, b) in the Figure, has performed relatively well, and gives a similar region to the Gnanadesikan-Kettenring correlation estimator at c) (both using $\mathrm{MAD_e}$ as scale estimator).  In this case, the MCD estimator, with \textit{h} set to 0.5 (not shown in the figure), gives a region almost identical to that from $\cov_{OGK}$ and identifies the same outliers.

 For this data set, the conclusions from the different outlier-resistant estimators are reasonably consistent: Laboratory 29 is an extreme case for all estimators; laboratories 27, 20 and 9 should be investigated; and laboratory 2 appears as a marginal 99\% outlier for two of the robust estimators and is barely inside the 99\% region from $\cov_{OGK}$. One other laboratory (not labelled in the plot) shows consistently as a 95\% outlier at the high end of the range and would normally be checked for technical issues, while a second in the same general area is just inside the 95\% region and would probably also merit further investigation.

There are small differences in detail between the different robust estimators. The Spearman-based estimator using $\mathrm{MAD_e}$ scale gave the smallest correlation and consequently the widest of the robust ellipses, perhaps reflecting the Spearman estimator's slight sensitivity to correlation outliers.   $\cov_{OGK}$ gave a high correlation resulting in a longer, narrow ellipse, probably due to the use of a scale estimator which is more efficient than $\mathrm{MAD_e}$ at the expense of somewhat greater univariate outlier sensitivity.  $\cov_{RGK}$ using $\mathrm{MAD_e}$ scale showed behaviour intermediate between $\cov_{OGK}$ and $\cov_{OGK}$, illustrating a modest improvement in resistance to correlation outliers compared to ${\cov_{RS}}$.    

Despite the expected small differences between different robust estimators, however, it is clear from this example that use of a robust region in a Youden plot provides considerably better discrimination of outlying values than the classical covariance.  

\section{Robust Mahalanobis distance}

A Mahalanobis distance is a covariance-scaled distance from (usually) a centroid to a particular point in a multivariate data set. It can be thought of as a multivariate analogue of the familiar \textit{z}-score, when the \textit{z}-score divisor is based on the dispersion of the data set. For a vector \textbf{z}$_{i}$, multivariate location ${\upmu}$ and a covariance matrix \textbf{V,} the Mahalanobis distance \textit{d}$_{\mathrm{M}}$ for \textbf{z}$_{i}$ is given by

\begin{equation}
d_{\mathrm{M}}=\sqrt{\left(\mathbf{z}_{i}- \mathbf{\upmu }\right)^{\mathrm{T}}\mathbf{V}^{- 1}\left(\mathbf{z}_{i}- \mathbf{\upmu }\right)}
\label{ref-005}
\end{equation}

In statistical software, Mahalanobis distance is often presented as a squared distance $d_{\mathrm{M}}^{2}$, that is, without taking the square root in eqn. \eqref{ref-005}. Lischer suggested its use in this form for proficiency testing \cite{Lischer1996}, as well as noting the idea of robust variants, though the low efficiency of available robust covariance methods at the time appears to have dictated another approach. When multivariate data with \textit{p} variables arise from a multivariate normal distribution with centroid ${\upmu}$ and covariance matrix \textbf{V,} $d_{\mathrm{M}}^{2}$ is distributed as ${\chi}^{2}$(\textit{p}), giving a simple approximate method of estimating critical values. Given robust estimators for covariance, a robust Mahalanobis distance (RMHD) can be calculated by combining a robust covariance matrix with a robust estimate of location. 

The application to inspection of interlaboratory data can be illustrated using the multi-element data set of {\hyperref[ref-039]{Table~S2}}. It is first necessary to decide how to treat the small number of missing values, which are common in inter-laboratory data and which would interfere with both covariance and distance calculations. For inspection purposes, a small proportion of missing values can be replaced with a suitable default value; here, missing values are replaced with the median of the remaining data in each column. This is a very simple form of `imputation'. For reliable statistical inference, more sophisticated processes are recommended, in particular multiple imputation methods \cite{mice2018}, but direct single replacement suffices for inspection.

\begin{figure}[t]
	\subfloat[]{%
		\includegraphics[width=0.47\textwidth]{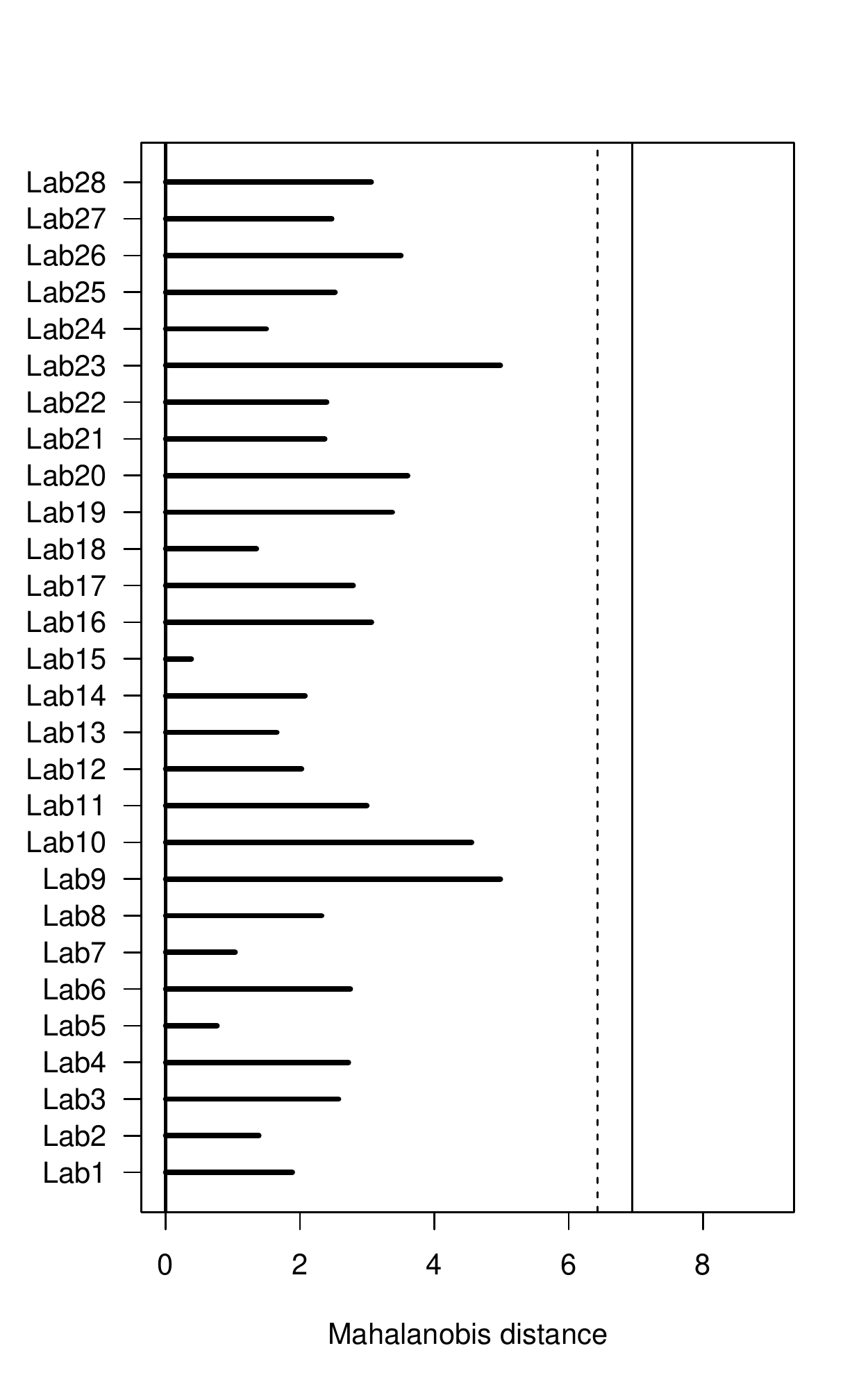} 
		\label{fig:mhd-a}}  
	\quad
	\subfloat[]{%
		\includegraphics[width=0.47\textwidth]{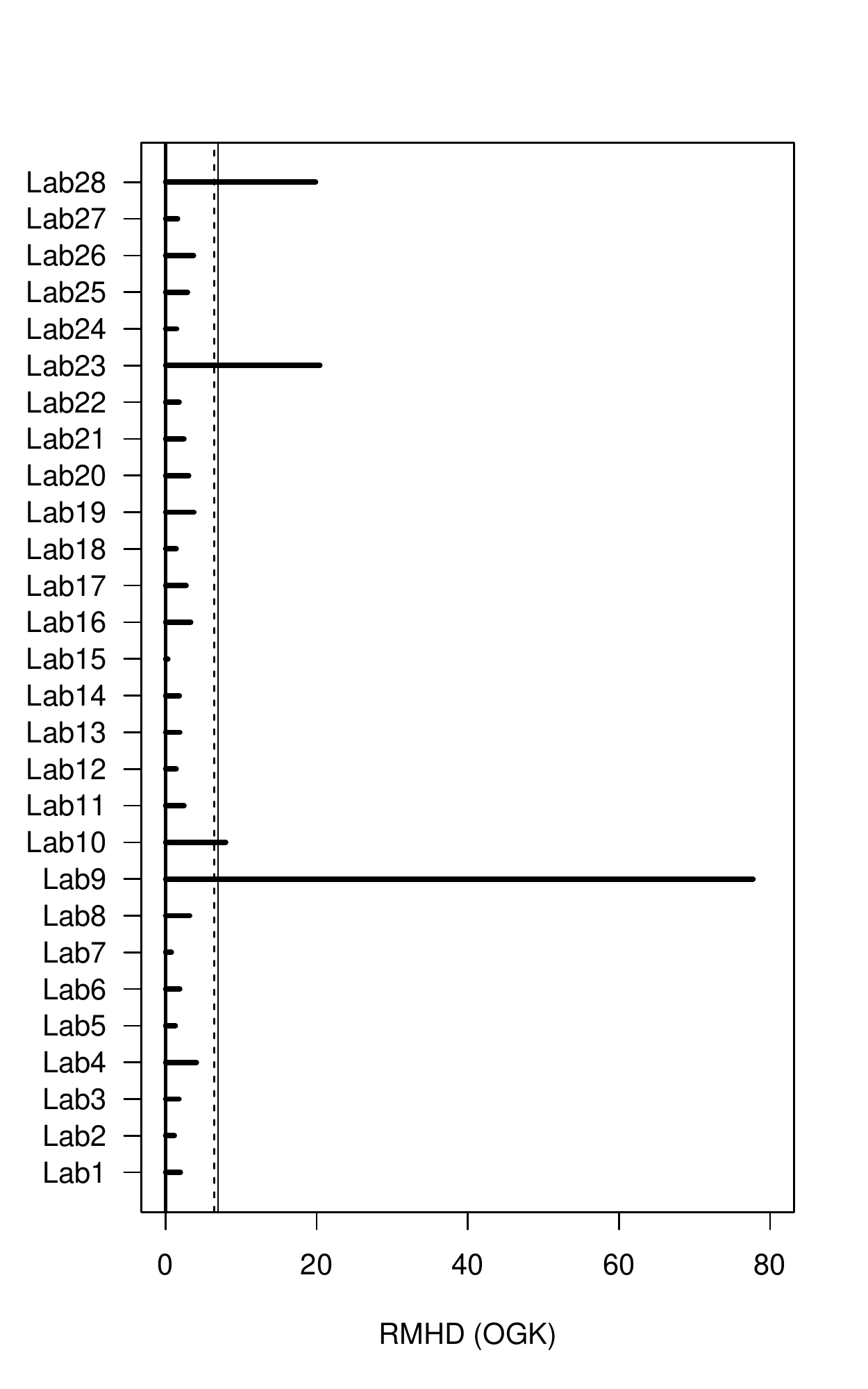} 
		\label{fig:mhd-b}} 
	\caption{Classical and robust Mahalanobis distances. \protect\subref{fig:mhd-a} classical Mahalanobis distance and \protect\subref{fig:mhd-b} Robust Mahalanobis distance (RMHD) calculated using the OGK estimate of covariance, for the data in {\hyperref[ref-039]{Table S2}} with simple imputation. Vertical lines show approximate critical values at the 95\% (dashed) and 99\% (solid line) levels of confidence.}
	\label{ref-036}
\end{figure}

Figure~\ref{ref-036} compares classical Mahalanobis distances for the data in {\hyperref[ref-039]{Table S2}} with the corresponding RMHD, based (in this case) on $\cov_{OGK}$. Distances are calculated from the data set means in Figure~\ref{ref-036}a) and from medians in Figure~\ref{ref-036}b); the use of medians for location in Figure~\ref{ref-036}a) made no difference to the general findings. The differences are striking. First, the classical variant in Figure~\ref{ref-036}a) shows no marked outlying laboratories. Second, none of the values in Figure~\ref{ref-036}a) is above either of the critical values shown in the plot. By contrast, the range of values is much larger in Figure~\ref{ref-036}b), and there are at least four laboratories which exceed the 99\% critical value, one very substantially so.

The interpretation of the RMHD plot in Figure~\ref{ref-036}b) is straightforward.  Laboratories 9, 23, 28 and, to a lesser extent, 10, are clearly anomalous and should be inspected in more detail. Inspection of the raw data shows, for example, that Laboratory 9 has a particularly high value for arsenic; laboratory 23 shows outlying values for cadmium and lead and reported no nickel in the sample; and laboratory 28 has reported unusually low values for both manganese and arsenic.

The reason for the poor discrimination shown by the classical estimator in Figure~\ref{ref-036}a) is primarily that multiple outliers for many elements have substantially inflated the covariance estimates, in turn reducing the calculated Mahalanobis distances to a point where they all appear to be within the expected 95\% region for a multivariate normal distribution. By contrast, use of a robust covariance estimator shows several important lines of investigation in a very simple plot.

It is, however, important to bear in mind that the Mahalanobis distance remains a summary statistic across multiple variables. While this can be very helpful in cases of multiple modest deviations leading to an unusual location (such as a modest correlation outlier), it is possible for any summary to conceal important detail. It follows that some additional detailed follow-up for individual, less striking, anomalies is usually prudent. 

\section{Robust principal component analysis}

Principal component analysis (PCA) is an important tool for exploratory analysis of multivariate data \cite{chatcol80}. It operates by choosing a set of orthogonal directions which successively represent most variation in the data set as a whole; choosing those directions associated with the largest variation effectively reduces the number of dimensions needed to identify structure in the data. PCA is, however, prone to interference from outlying values in the same way as most other statistical methods;  extreme values can dominate variation and distort the principal components. This can, however, be overcome by use of robust estimators for covariance and correlation. This follows because, although PCA can be performed in a number of ways, all the principal components for a given data set can be obtained directly from a correlation or covariance matrix by eigenvector decomposition \cite{chatcol80}. (The use of correlation and covariance in PCA are not equivalent; PCA from a correlation matrix is essentially equivalent to PCA on \textit{z}-scaled data. This gives different principal components. PCA based on correlation is often recommended where variables differ considerably in magnitude). Given a method of obtaining robust estimates of covariance, together with standard tools for eigenvector decomposition, all that is needed to obtain robust principal components is to replace the classical covariance matrix with the robust covariance matrix in the PCA extraction step. Eigenvector extraction is not straightforward, and is best handled by specialised routines; it is not discussed here. Fortunately, reliable implementations are widely available as standard software libraries and in both open source and commercial mathematical and statistical software, making application to inter-laboratory data almost as straightforward as classical PCA.

\begin{figure}[!h]
	\centering
	\subfloat[]{%
		\includegraphics[width=0.47\textwidth]{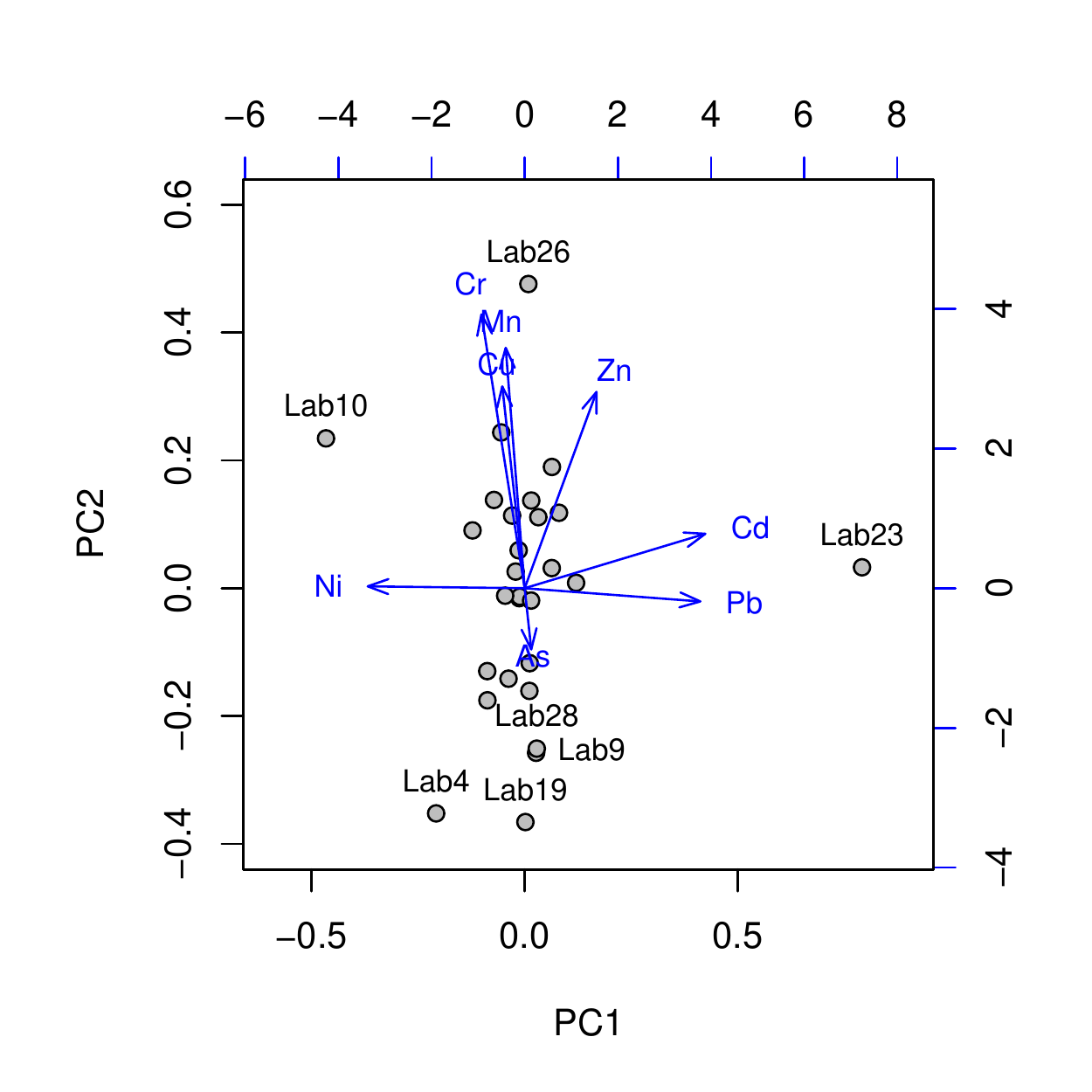}
		\label{fig:pca-a} 
	} 
	\quad
	\subfloat[]{%
		\includegraphics[width=0.47\textwidth]{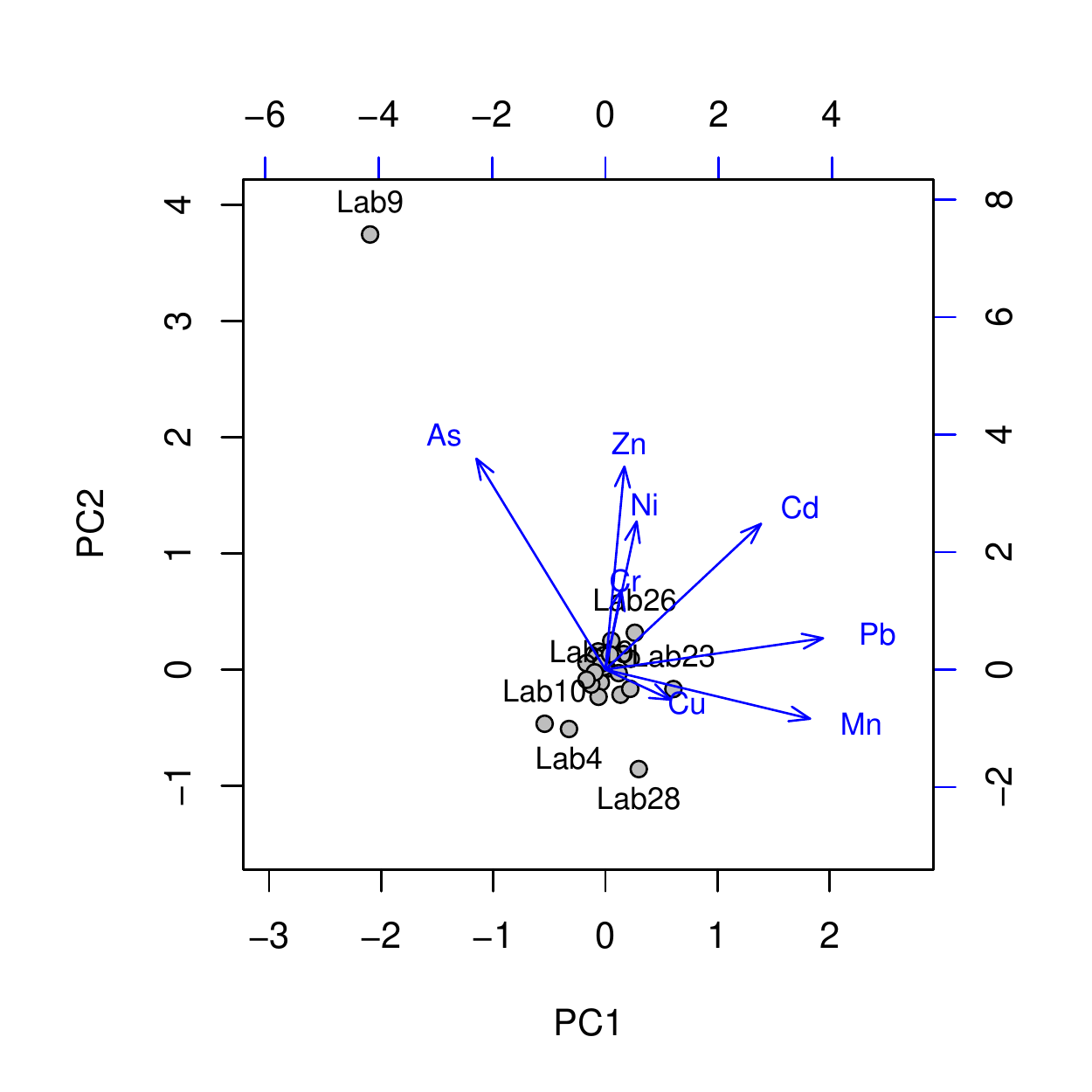}
		\label{fig:pca-b} 
	} 
	\caption{Classical and robust PCA for inter-laboratory data.  \protect\subref{fig:pca-a} Classical PCA biplot for the first two principal components for the data in Table~S2 constructed from the classical covariance matrix; \protect\subref{fig:pca-b} Biplot for PCA based on a covariance matrix constructed using the MCD estimator. Variables were scaled in both cases. Points (filled circles) show scores on PC1 and PC2 for each laboratory; arrows show the projections of the different variables on the PC1/PC2 plane. Only the outlying laboratories in both plots are labelled.}

	\label{ref-037}
\end{figure}

The utility of robust PCA can be illustrated by application to the data of {\hyperref[ref-039]{Table S2}}. Figure~\ref{ref-037} compares biplots for the first two principal components obtained from classical (Figure~\ref{ref-037}a) and robust (Figure~\ref{ref-037}b) PCA. The robust method used the MCD estimator $\cov_{MCD}$ for covariance; because of the wide range of concentrations, both analyses used scaling by variable.

The most striking feature is the difference in apparent dispersion among laboratories. For the classical PCA, laboratories are relatively dispersed, especially along PC2. In the robust plot, the distance to laboratory 9 dwarfs other distances in the diagram, with laboratories 23 and 28 visible at the bottom of the diagram. There are also differences in detail. In the classical PCA plot, laboratories 23 and, to a lesser extent, 26 and 10 appear most distinct. From the biplot, Lab 23 is apparently associated with high cadmium and lead values (as noted above) and low nickel; Lab 26 with high chromium, copper and manganese and low arsenic; Lab 10 approximately mirrors the associations of Lab 23. Lab 9 is not prominent. By comparison, the robust PCA biplot of Figure~\ref{ref-037}b) clearly highlights Lab 9 as a considerable outlier, with Lab 23 and Lab 28 again apparently distinct. Expansion of the central region of Figure~\ref{ref-037}b) (not shown) identifies Lab 10 and (to a lesser extent) Lab 4 as modest outlying points to low left of the majority. This clearly follows the indications given by the robust Mahalanobis plot (Figure~\ref{ref-036}); the important additional information is an indication of the elements likely to be responsible for the anomalous locations. For example, Lab 9's location is apparently associated with very high arsenic. This partially explains why Lab 9 is not obvious in Figure~\ref{ref-037}a), as arsenic does not contribute strongly to the first two classical principal components. 

A further difference between the classical and robust PCA is the range of principal component scores, which indicate the relative distance, from a centroid, of a given point along each principal component direction. The classical PCA shows a small range of scores along the first two PCs, which (like the apparently small classical Mahalanobis distances in Figure~\ref{ref-036}a) might be taken to indicate that no values are particularly important. By comparison, scores in the robust PCA show a substantially larger range, more accurately indicating that the outliers are more remote from a central majority than might be expected by chance. Thus, while systematic exploration through different pairwise choices of principal component can generally locate outliers regardless of the use of robust methods, extreme cases appear to stand out more strongly, with more informative PC scores, in robust PCA.

\section{Conclusions}

This paper has briefly drawn attention to a number of robust estimators of covariance and provided illustrations of some applications of robust covariance in the examination of inter-laboratory data, exemplified by a reference material certification exercise. Robust covariance estimators are much less affected by outlying values, in turn providing more reliable estimates of covariance in outlier-contaminated data sets. As a consequence, the use of robust covariance estimators led to substantially clearer diagnostics; anomalies were more readily detected on visual inspection, and critical values based on robust estimates of (co)variance identify important extremes more reliably. There is therefore good reason to include robust covariance estimators, and associated diagnostics, in the inter-laboratory toolkit.

Turning to the most appropriate robust estimator for inspection of inter-laboratory data, among those explored here, the Orthogonalized Gnanadesikan-Kettenring (OGK) estimator seems most generally applicable when a suitable software implementation is available. It provides broadly similar performance to the MCD estimator, offers fewer restrictions on number of variables, and guarantees a positive definite covariance matrix. It also offers a speed advantage over MCD, though in practice this is inconsequential for typical interlaboratory data sets. In the absence of a software implementation of OGK or MCD estimators, the much simpler $\cov_{RGK}$ estimator is a good alternative, particularly for pairwise correlation and covariance; it has useful resistance to correlation outliers and guarantees valid pairwise correlation. In the rare cases where multiple pairwise application of $\cov_{RGK}$ generates an invalid multivariate covariance matrix, the combination of rank correlation and robust standard deviation guarantees non-singular covariance matrix at the expense of somewhat greater susceptibility to off-diagonal ``correlation outliers''. 

\section*{Acknowledgements}
Preparation of this paper was supported by the UK Department for
Business, Energy and Industrial Strategy under the 
UK National Measurement System Chemical and Biological
Metrology Programme. 

The author is additionally grateful to Professor M Thompson (Birkbeck college, London) for helpful comments on an early draft of this manuscript.

\section*{Conflicts of interest}
There are no conflicts of interest to declare.

\bibliographystyle{ieeetr}
\bibliography{robcov}

\clearpage
\section*{Annex: Stepwise construction of a data ellipse}

Plotting coordinates for a robust data ellipse suitable for a Youden plot can be constructed from the (robust) means  $\bar{x}_i^*$, $\bar{x}_j^*$, robust standard deviations $\hat{s}_i$ and $\hat{s}_j$, and robust covariance $\cov_{ij}^*$ as shown below. The construction shown here generates the upper and lower portions of an ellipse separately and concatenates the coordinates; for calculation in a spreadsheet, the lower set of coordinates (using intermediate values with subscript ``-'') are simply calculated in rows immediately below the calculation for the upper set, denoted ``+''. The general form of the construction follows Jackson \cite{jacksonIQC}. 
\begin{enumerate}[i)]
	\item Calculate the robust correlation coefficient $r_{ij}^*$ from  $\hat{s}_i$, $\hat{s}_j$, and $\cov_{ij}^*$ using equation \eqref{ref-002};
	
	\item Calculate $T^2$ using 
	$$T^2 = 2\left(n - 1 \right) f\left(p, 2, n-1\right)/(n-2)$$
	where $n$ is the number of observations used to calculate the robust covariance, $p$ is the desired coverage probability (for example, 95\%) and$f(p, 2, n-1)$ is the upper $p$ quantile for the F distribution with 2 and $(n-1)$ degrees of freedom. (This is implemented as \texttt{F.INV($p$,2,$n-1$)} in MS Excel);
	
	\item Let $\theta$ be a series of $n_{\rm points}$ angles from $\pi$ to zero radians  (inclusive), $n_{\rm points}$ being chosen to give a reasonably smooth curve on plotting (for example, $n_{\rm points}=100$). Calculate the $n_{\rm points}$ coordinates $z_{x+, k}$, $k=1, \ldots, n_{\rm points}$
	$$z_{x+, k} = T \cos(\theta_{k}), \, k=1, \ldots, n_{\rm points}$$
	and set the additional coordinates $z_{x-}$ (in a spreadsheet, in the rows below  $z_{x+}$) to
	$$z_{x-, k} = T \cos(\theta_{k}), \, k=n_{\rm points}-1, \ldots, 2$$
	 $z_{x-}$ then contains the values from $z_{x+}$ in reverse order, omitting the points for $\theta=0$ and $\theta=\pi$. For spreadsheet calculation, it may be useful to add a second set of angular coordinates $\theta_{-}$ running from 0 to $\pi$, omitting the values for $\theta=0$ and $\theta=\pi$, and calculate $z_{x-}$ from the corresponding angular coordinates.
	
	\item Calculate upper and lower coordinates $z_{y+}$ and  $z_{y-}$ from
	$$z_{y+, k} = r_{ij}^* z_{x+, k} + \sqrt{\left[1-(r_{ij}^*)^2\right]\left(T^2-z_{x+, k}^2\right)}, \,  k=1, \ldots, n_{\rm points}$$
	$$z_{y-, k} = r_{ij}^* z_{x-, k} - \sqrt{\left[1-(r_{ij}^*)^2\right]\left(T^2-z_{x-, k}^2\right)}, \,  k=1, \ldots, n_{\rm points}-2$$
	The second calculation runs over two fewer points as $z_{x-, k}$ is a shorter sequence than $z_{x+, k}$. In a spreadsheet, the points $z_{y}$ can conveniently be calculated in the column adjacent to the points $z_{x}$. Where the term inside the square root becomes adventitiously negative due to rounding, it may be set to zero.

	\item Taking $z_x$ as the concatenation of $z_{x+}$ and  $z_{x-}$, and  $z_y$ the corresponding points from  $z_{y+}$ and  $z_{y-}$, calculate plotting coordinates $(x_p, y_p)$ as
	$$x_{p,k} = \hat{s}_i z_{x,k} + \bar{x}_i^*, \, k=1, \ldots, 2n_{\rm points}-2$$
	$$y_{p,k} = \hat{s}_j z_{y,k} + \bar{x}_j^*, \, k=1, \ldots, 2n_{\rm points}-2$$

\end{enumerate}
Notes: 
\begin{enumerate}[a)]
	\item The points comprise a polygon to be plotted as a closed curve. Where a line plot is used (as in most spreadsheets), the first point should be repeated at the end of the data set to provide a closed curve. 
	
	\item When provided with a robust covariance matrix $\mathbf{V}$, $\hat{s}_i$ and $\hat{s}_j$ are the square root of the corresponding diagonal elements $V_{ii}$ and  $V_{jj}$, and $\cov_{ij}^*$ is the off-diagonal element $V_{ij}$.
	
	\item If the covariance and standard deviations are assumed to be known 
	(that is, taken as population values), $T^2$ can be calculated using 
	the chi-squared distribution with 2 degrees of freedom using
	$$T^2 = \chi_{p,2}^2 /2 $$
	where $\chi_{p,2}^2$ is the upper $p$ quantile for the $\chi^2$ distribution with 2 degrees of freedom. 
\end{enumerate}

\clearpage
\section*{Supplementary material}

\begin{table}[h]
	\caption*{Table S1: Potassium data from an inter-laboratory certification exercise}
	\centering
	\begin{tabular}{c c c}
	\hline 
	
	\textbf{Laboratory~ID} & \textbf{QC} & \textbf{RM} \\
	\hline 
	Lab01 & 7.9367 & 5.1640 \\
	Lab02 & 9.3400 & 5.9400 \\
	Lab03 & 7.3969 & 4.7404 \\
	Lab04 & 7.6350 & 5.1580 \\
	Lab05 & 7.6700 & 4.9720 \\
	Lab06 & 8.2500 & 5.4080 \\
	Lab07 & 7.7600 & 5.0840 \\
	Lab08 & 8.2700 & 5.1900 \\
	Lab09 & 10.1200 & 6.5580 \\
	Lab11 & 7.9900 & 5.1620 \\
	Lab12 & 7.9300 & 5.0980 \\
	Lab13 & 8.7933 & 5.7520 \\
	Lab14 & 7.8533 & 4.9440 \\
	Lab16 & 7.8500 & 5.4060 \\
	Lab18 & 7.6600 & 4.7000 \\
	Lab19 & 7.7800 & 5.1800 \\
	Lab20 & 9.0600 & 5.1960 \\
	Lab21 & 7.6191 & 4.9121 \\
	Lab22 & 7.4167 & 4.7480 \\
	Lab23 & 8.1000 & 5.2800 \\
	Lab25 & 7.8700 & 5.1660 \\
	Lab26 & 9.0858 & 5.7634 \\
	Lab27 & 6.7433 & 3.8200 \\
	Lab28 & 7.8167 & 4.9400 \\
	Lab29 & 5.2550 & 7.7900 \\
	\hline 
	\end{tabular}
	\label{ref-038}
	
	\caption*{The table gives results obtained for potassium in an inter-laboratory reference material certification exercise. Two materials were circulated; a candidate drinking water reference material (``RM'') and a performance control material (``QC'') made up by spiking demineralised water. All values (shown to four decimal places) are in mg~L$^{-1}$. Laboratories measured multiple replicates of each material; the data set comprises the means for each laboratory.}

\end{table}

\begin{landscape}
	
\begin{table}[p]

\caption*{Table S2: Eight elements from a reference material certification study}
\centering
\begin{tabular}{l x{1.5cm}  x{2.1cm}  x{1.5cm}  x{1.5cm}  x{1.5cm}  x{2cm}  x{1.5cm}  x{1.5cm}}
\hline 
\textbf{Laboratory ID} 
& \textbf{Arsenic} ${\mu\mathrm{g~L}^{-1}}$ 
& \textbf{Cadmium}  ${\mu\mathrm{g~L}^{-1}}$ 
& \textbf{Chromium}  ${\mu\mathrm{g~L}^{-1}}$
& \textbf{Copper}  ${\mu\mathrm{g~L}^{-1}}$
& \textbf{Lead}  ${\mu\mathrm{g~L}^{-1}}$
& \textbf{Manganese}  ${\mu\mathrm{g~L}^{-1}}$
& \textbf{Nickel}  ${\mu\mathrm{g~L}^{-1}}$
& \textbf{Zinc}  ${\mu\mathrm{g~L}^{-1}}$ \\
\hline 
Lab1 & 10.014 & 5.0900 & 48.084 & 2016.0 & 25.290 & 50.632 & 19.740 & 613.44 \\
Lab2 & 10.288 & 4.9880 & 48.166 & 1936.4 & 24.240 & 47.246 & 19.214 & 634.52 \\
Lab3 & 10.166 & 4.9719 & 47.373 & 1682.4 & 22.893 & 48.073 & 18.525 & 598.21 \\
Lab4 & 9.096 & 4.4700 & 44.382 & 1882.2 & 21.202 & 44.310 & 19.568 & 551.14 \\
Lab5 & 10.004 & 4.8880 & 49.654 & 1970.8 & 23.976 & 48.100 & 19.624 & 607.57 \\
Lab6 & 10.440 & 4.9560 & 49.820 & 1900.0 & 22.560 & 48.700 & 18.520 & 654.20 \\
Lab7 & 10.342 & 4.9220 & 50.368 & 1938.3 & 23.256 & 49.020 & 19.944 & 620.32 \\
Lab8 & 10.474 & 4.8440 & 45.712 & 2068.2 & 23.670 & 46.668 & 20.616 & 625.04 \\
Lab9 & 30.916 & 4.6120 & 44.742 & 1959.9 & 26.592 & 47.654 & 20.340 & 582.48 \\
Lab10 & 10.120 & 3.9580 & 54.480 & 2048.0 & 19.060 & 51.620 & NA & 578.00 \\
Lab11 & 10.700 & 4.9800 & 48.540 & 2013.8 & 26.520 & 45.300 & 19.880 & 626.06 \\
Lab12 & 9.876 & 4.8220 & 46.086 & 1827.1 & 23.780 & 48.072 & 18.740 & 608.12 \\
Lab13 & 10.440 & 5.1020 & 51.160 & 2026.0 & 24.920 & 50.500 & 18.340 & 603.20 \\
Lab14 & 10.412 & 4.8580 & 49.300 & 1845.2 & 22.494 & 47.460 & 18.288 & 554.12 \\
Lab15 & 10.200 & 4.8960 & 48.960 & 1956.2 & NA & 48.376 & 19.528 & NA \\
Lab16 & 9.603 & 4.9120 & 47.108 & 2225.2 & 24.710 & 49.298 & 17.432 & 592.26 \\
Lab17 & 9.934 & 4.8220 & 50.520 & 2096.0 & 22.260 & 49.700 & 17.580 & 556.80 \\
Lab18 & 10.502 & 4.8616 & 47.556 & 1830.5 & 22.870 & 46.106 & 19.528 & 597.31 \\
Lab19 & 9.942 & 4.8120 & 47.182 & 1686.8 & 24.706 & 43.656 & 19.184 & 556.13 \\
Lab20 & 9.534 & 4.9180 & 47.916 & 1803.7 & 24.950 & 53.564 & 20.032 & 564.96 \\
Lab21 & 10.362 & 4.7042 & 51.594 & 2054.7 & 24.047 & 50.012 & 19.823 & 633.50 \\
Lab22 & 10.314 & 4.9660 & 52.684 & 1938.2 & 23.550 & 49.774 & 20.840 & 586.24 \\
Lab23 & NA & 6.0000 & 48.200 & 1886.0 & 30.000 & 47.800 & 0.000 & 620.60 \\
Lab24 & 10.180 & 4.8520 & 47.740 & 2040.0 & 23.040 & 46.720 & 19.320 & NA \\
Lab25 & 10.054 & 4.9660 & 46.266 & 1893.2 & 24.262 & 51.548 & 19.628 & 590.62 \\
Lab26 & 9.794 & 5.2200 & 55.467 & 2019.0 & 23.160 & 51.945 & 21.162 & 663.69 \\
Lab27 & NA & NA & NA & 1865.2 & 22.026 & 45.982 & 18.806 & 559.93 \\
Lab28 & 5.342 & NA & 45.660 & 1906.4 & NA & 40.862 & NA & 607.51 \\
Lab29 & 12.420 & 6.0300 & 55.033 & 1888.7 & 30.013 & 50.173 & 19.977 & 589.88 \\
\hline 
\end{tabular}
\label{ref-039}
\caption*{The table shows the mean of (nominally) five replicates per laboratory for each of eight elements in a reference material certification exercise. NA denotes missing values; that is, the laboratory did not report results for the element concerned.}
\end{table}

\end{landscape}

\end{document}